\documentclass[prl,twocolumn,showpacs,preprintnumbers,amsmath,amssymb]{revtex4}
\usepackage{graphicx}% Include figure files
\usepackage{dcolumn}% Align table columns on decimal point
\usepackage{bm}% bold math

\begin{document}

\input{epsf}
%\preprint{}

\title{Spin waves in a Bose Ferromagnet}

\author{Qiang Gu, Kai Bongs and Klaus Sengstock}

\affiliation{Institut f\"ur Laser-Physik, Universit\"at Hamburg,
Luruper Chaussee 149, 22761 Hamburg, Germany}

\date{\today}
\textbf{}

\begin{abstract}
It is shown that the ferromagnetic transition takes place
always above Bose-Einstein condensation in ferromagnetically
coupled spinor Bose gases. We describe the Bose ferromagnet within
Ginzburg-Landau theory by a ``two-fluid" model below Bose-Einstein
condensation. Both the Bose condensate and the normal phase are
spontaneously magnetized. As a main result we show that spin waves
in the two fluids are coupled together so as to produce only one
mixed spin-wave mode in the coexisting state.
The long wavelength spectrum is quadratic in the
wave vector ${\bf k}$, consistent with usual ferromagnetism
theory, and the spin-wave stiffness coefficient $c_s$ includes
contributions from both the two phases, implying the ``two-fluid"
feature of the system. $c_s$ can show a sharp bend at the
Bose-Einstein condensation temperature.
\end{abstract}


\maketitle

Ferromagnetism belongs to the oldest phenomena in condensed
matters and it is attracting continuous research interest till
today\cite{mohn}, sparked by discoveries of new materials or new
phenomena in this field. Most ferromagnets being studied so far
are composed of fermionic particles. The recent experimental
success of optical confining ultracold atomic Bose
gases\cite{ketterle,barrett} provides possibilities to realize a
new kind of ferromagnet, the Bose ferromagnet. A promising example
might be the gas of $F=1$ $\rm ^{87}Rb$ atoms. It has been
predicted in theory\cite{burke} and confirmed by
experiments\cite{schma,chang} that the hyperfine spin-spin
interaction in this Bose gas is ferromagnetic.

In dilute atomic gases, interatomic forces are rather weak, with
the effective $s$-wave scattering length being typically of the
order 100$a_B$ where $a_B$ is the Bohr radius. The
spin-dependent interaction is even 1 or 2 orders of magnitude
smaller\cite{burke}. One may question whether the ferromagnetic
(FM) transition induced by such a weak FM coupling could be
observed in experiments. Recently, it was already shown that
the FM transition in Bose gases takes place always above
Bose-Einstein condensation (BEC), regardless of the value of the
FM coupling\cite{Gu2}. It means that Bose gases can exhibit
ferromagnetism at relatively high temperatures in comparison to
the energy scale of the FM coupling. Below BEC, the FM Bose gas
becomes a ``two-fluid" system: the polarized Bose condensate coexisting
with the magnetized normal gas. This is a unique feature in the
Bose ferromagnet.

Accompanying the FM transition, spin waves appear in the
ferromagnet as a Goldstone mode. In conventional ferromagnets,
both insulating and metallic, the spin-wave excitations are
gapless at wave vector ${\bf k}=0$ and the long wavelength
dispersion relation is quadratic in ${\bf k}$, $\omega_s=c_s k^2$,
with the spin-wave stiffness coefficient $c_s$ proportional to the
strength of the FM coupling\cite{mohn}. It is a rather interesting
problem how the spin wave manifests itself in the Bose
ferromagnet, especially considering the ``two-fluid" feature of
the system. In this letter, we shall examine the phase diagram and
spin waves in the Bose ferromagnet phenomenologically and show how
spin waves in the two fluids couple together.

\begin{figure}
\center{\epsfxsize=75mm \epsfysize=40mm \epsfbox{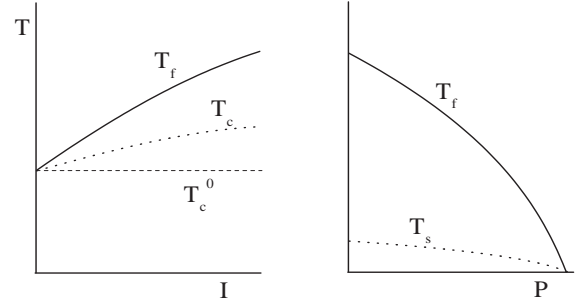}}
\caption{ \label{fig:epsart} Schematic Phase diagram of the Bose
ferromagnet (left) and the SC ferromagnet $\rm ZrZn_2$ (right). In
the left figure, $T_f$ is the FM transition temperature. $T_c$ and
$T^0_c$ denote the BEC temperatures for the Bose gas with and
without FM couplings (denoted as $I$) respectively. In the right
figure, $T_f$ and $T_{s}$ are the FM and SC transition
temperatures, respectively. The pressure $P$ is supposed to modify
the effective spin-exchanges between electrons in SC
ferromagnets.}
\end{figure}

Spin waves in spinor Bose-Einstein condensates were intensively
studied in theory\cite{ho,ohmi,ueda,szepfalusy}, motivated by
experiments on spinor BEC\cite{ketterle}. In their pioneering
papers, Ho\cite{ho} and Ohmi and Machida\cite{ohmi} calculated
low-lying collective modes both in ferro- and in antiferro-magnetic
condensates. For the FM case, a Bogliubov spectrum for the
density fluctuation and a $k^2$-formed dispersion for the spin
fluctuation were derived,
\begin{equation}
\hbar\omega_1 = \sqrt{\epsilon_k^2 + 2(g_0+g_s)n_0\epsilon_k},~~~~
\hbar\omega_0 = \epsilon_k = \frac {\hbar^2k^2}{2m} ,
\end{equation}
coinciding with available theories of BEC and ferromagnetism. Here
$g_0$ and $g_s$ denote the spin-independent and dependent
interactions, respectively. A surprise in their results is that
{\it the spin-wave stiffness, $c_s={\hbar^2}/(2m)$, does not depend
on the FM interaction}. The same results were obtained by Ueda
through a many-body mean-field approach\cite{ueda}. Also many
attempts were made to generalize these results to finite
temperatures\cite{szepfalusy}. However, all these works are
dedicated to only the Bose condensed phase, without taking into
account the magnetized normal phase.

Analogous with the Bose ferromagnet, coexistence of ferromagnetism
and superconductivity was discovered in various solid state
materials\cite{saxena}, such as $\rm UGe_2$, $\rm ZrZn_2$ and $\rm
URhGe$. It is observed that these materials first undergo a FM
transition, then go into the superconducting (SC) state. Their
temperature-pressure phase diagram seems quite similar to the
temperature-FM coupling phase diagram of the Bose
ferromagnet\cite{Gu2}, as shown in Figure 1. In such SC ferromagnets,
both ferromagnetism and superconductivity are believed due to the
same band electrons and the Cooper pairs are most likely in the
triplet configuration\cite{saxena}. Thus the system behaves like a
spin-1 Bose gas to some extent.

Phenomenological theory is proved to be a powerful tool to
investigate behaviors of superconductors and superfluids near the
transition temperature. It has already been successfully employed
to describe various phase transitions\cite{machida} and to explain
the phase diagram\cite{walker,shopova} of the above-mentioned SC
ferromagnets, and also has ever been applied to spinor Bose
gases\cite{siggia,Gu1}. Although phenomenological theory only
produces qualitative results, it leads to deep insight into the
nature of the associated phenomenon.

We start with an appropriate Ginzburg-Landau (GL) free energy
density functional allowing for the coexistence of BEC and
ferromagnetism. It consists of three different parts: $f_t = f_m +
f_b + f_c$, with
\begin{subequations}
\begin{eqnarray}
f_m&=& c |\nabla {\bf M}|^2
    + \frac a2{|{\bf M}|^2} +  \frac b4 {|{\bf M}|^4} ,\\
f_b&=& \frac {\hbar^2}{2m} \nabla {\bf \Psi}^{\dag} \cdot \nabla {\bf \Psi}
    + \alpha |{\bf \Psi}|^2 + \frac {\beta_0}2 |{\bf \Psi}|^4
      \nonumber\\
   && + \frac {\beta_s}2 \Psi^*_\sigma \Psi^*_{\sigma'}
      {\bf F}_{\sigma\gamma} \cdot {\bf F}_{\sigma'\gamma'}
      \Psi_{\gamma'} \Psi_{\gamma}  ~,\\
f_c&=& -g {\bf M} \cdot \Psi^*_{\sigma}{\bf F}_{\sigma\gamma}
      \Psi_{\gamma} ~,
\end{eqnarray}
\end{subequations}
where repeated sub-indices represent summation taken over all the
hyperfine states. This model is based on the two-fluid description
of the Bose ferromagnet. $f_m({\bf M})$ describes the
ferromagnetic normal phase, which is the standard Landau free
energy density for the second-order FM transition, having the
magnetization density ${\bf M}$ as the order
parameter\cite{note2}. $f_b({\bf \Psi}^{\dag},{\bf \Psi})$
describes the Bose condensed phase. ${\bf \Psi}^{\dag}\equiv ({\bf
\Psi}^T)^{*}$ is the complex order parameter of the spinor
condensate. $m$ is the mass of the particle. $c$, $a$, $b$,
$\alpha$, $\beta_0$ and $\beta_s$ are phenomenological parameters.
$c$ is the spin-wave
stiffness in the FM normal gas. $a$ and $\alpha$ depend on the
temperature, $a=a^\prime(T-T_f)$ and
$\alpha=\alpha^\prime(T-T_c^0)$ where $T_f$  and $T_c^0$ are the
FM transition and BEC temperatures in the {\it decoupled case},
respectively. $\alpha^\prime$, $a^\prime$, $b$ and $\beta_0$ are
positive constants and independent of the temperature. The
$\beta_s$ term comes from the spin-dependent interaction and thus
has SO(3) symmetry. ${\bf F}=(F_x, F_y, F_z)$ are Pauli matrices
for the hyperfine spin. $\beta_{s}<0$ in a Bose ferromagnet. $f_c$
describes the coupling between the two phases, with the coupling
constant $g>0$.

The full FM order parameter in the normal gas takes the form ${\bf
M} =\langle {\bf M} \rangle+\delta{\bf M}$. Its average value is
chosen to be along the $z$ direction for convenience, $\langle
{\bf M} \rangle =(0,0,M_0)$. $\delta{\bf M}=(\delta M_x,\delta
M_y,\delta M_z)$ represents spin fluctuations. The full order
parameter for the condensate is written as ${\bf \Psi}^{\dag} =
(\Phi_1^{*}+\delta \Psi_1^{*}, \Phi_0^{*}+\delta \Psi_0^{*},
\Phi_{-1}^{*}+\delta \Psi_{-1}^{*})$. $\Phi_\sigma = \langle
\Psi_\sigma \rangle$ and $\Phi_\sigma^{\dag}\Phi_\sigma=n_0$ is
just the density of condensed bosons. For a homogeneous system,
both $M_0$ and $\Phi_\sigma$ are spatially independent, so $\nabla
M_0 = \nabla \Phi_\sigma = 0$.

First, we examine the phase diagram predicted from a microscopic
model in Ref. \protect\cite{Gu2}. Neglecting all the fluctuations
and minimizing the total free energy density $f_t$ with respect to
$\Phi^\dag$, one gets the stable solution for the condensate. It
is found that only spin-$1$ bosons condense,
\begin{eqnarray}
|\Phi_1|^2 = \frac {\alpha^\prime}{\beta_0+\beta_s} \left(
T-T_c^0-\frac g{\alpha^\prime}M_0 \right) ,
\end{eqnarray}
and $|\Phi_0|^2 = |\Phi_{-1}|^2 = 0$. These results agree with the
prediction of Ho\cite{ho} and Ohmi and Machida\cite{ohmi}, and
also the microscopic theory\cite{Gu2}. Obviously, the BEC critical
temperature is enhanced by $M_0$, the magnetization in the normal phase,
\begin{eqnarray}\label{eq22}
T_c=T_c^0+\frac{g}{\alpha^\prime}M_0 .
\end{eqnarray}
At $T=T_c$, the order parameter of the condensate is zero, and we
can derive the value of $M_0$ by minimizing $f_m(M_0)$ with respect
to $M_0$,
\begin{eqnarray}\label{eq23}
M_0 = \sqrt{\frac {a^\prime}{b}[T_f-T_c]}.
\end{eqnarray}
Combining Eq. (\ref{eq22}) and (\ref{eq23}), one has
$\delta T_c=\sqrt{ T^* (\delta T_f-\delta T_c)}$,
with $T^*=(g/\alpha^\prime)^2a^\prime/b$, $\delta T_c=T_c-T_c^0$
and $\delta T_f=T_f-T_c^0$. At very small ferromagnetic couplings,
$I\to 0$, both $\delta T_c$ and $\delta T_f$ tend to zero with $I$.
Provided that $\delta T_f<<T^*$\cite{walker}, we have
\begin{eqnarray}\label{eq24}
(\delta T_c)\approx \delta T_f\left(1-\frac {\delta T_f}{T^*}\right) .
\end{eqnarray}
Suppose $\delta T_f$ depends linearly on $I$ when $I<<1$\cite{note3},
$\delta T_f=CI$, the $I$-dependence of $\delta T_c$ is given by
\begin{eqnarray}\label{eq25}
\delta T_c=CI\left(1-\frac {CI}{T^*}\right) .
\end{eqnarray}
Besides a linear term, $\delta T_c$ also depends on $I$
quadratically. So far, the phase diagram of Bose ferromagnets is
roughly reproduced. Similar treatment had been carried out to
explain the $T-P$ phase diagram of SC ferromagnets
previously\cite{walker,shopova}.

It was argued that BEC itself cause a FM transition in spinor
bosons\cite{yamada}. If one considers only the free energy density
of the Bose condensate, the $\beta_s$ term alone indeed can break
the $SO(3)$ symmetry\cite{ho,ohmi}. However, since FM transition
occurs already above BEC, the $\beta_s$ term no longer plays a
critical role. It only enhances the condensate fraction, but does
not result in another FM transition. This point is confirmed by the
microscopic theory\cite{Gu2}, which shows that both the
magnetization $M$ and the derivatives $(\partial M/\partial T)$
seem smooth at $T_c$.

To proceed, we discuss spin waves in the Bose ferromagnet. Within the
GL theory, the spin wave is interpreted as spin fluctuations
around the average value of the order parameter. To linear order
in $\delta{\bf M}$ and $\delta\Psi_\sigma$, the fluctuations are
described by $f^I_t=f^I_m+f^I_b+f^I_c$, with
\begin{subequations}\label{flu}
\begin{eqnarray}
f^I_m &=& c \nabla {\delta M_+} \nabla {\delta M_-}
      + \frac 12 \left(a+bM_0^2\right) {\delta M_+}{\delta M_-} ,\\
f^I_b &=& \frac {\hbar^2}{2m} \nabla{\delta\Psi^*_\sigma}
    \nabla {\delta\Psi_\sigma} + [\alpha+ (\beta_0+\beta_s)
    {\Phi_1}^2] {\delta\Psi^*_\sigma} {\delta\Psi_\sigma}
  \nonumber\\
  &&+ \frac 12 (\beta_0+\beta_s) {\Phi_1}^2 ({\delta\Psi^*_1}
    + {\delta\Psi_1})^2
  \nonumber\\
  &&- 2{\beta_s}{\Phi_1}^2 {\delta\Psi^*_{-1}}{\delta\Psi_{-1}} ,\\
f^I_c &=& -g M_0(\delta\Psi^*_1\delta\Psi_1
    - \delta\Psi^*_{-1}\delta\Psi_{-1})
  \nonumber\\
  &&-\frac {\sqrt{2}}2 g \Phi_1 (\delta\Psi^*_0{\delta M_+}
    + \delta\Psi_0{\delta M_-}).
\end{eqnarray}
\end{subequations}
Here $\delta M_+=\delta M_x+i\delta M_y$ and $\delta M_+=\delta
M_x-i\delta M_y$ describe the transverse spin fluctuations (i.e.
the magnons) in the normal gas. The longitudinal spin fluctuation
is neglected because it is a gapped mode in the FM phase and can
hardly be excited at low energy\cite{mohn}. As pointed out in
{Refs.\ \protect\cite{ho,ohmi}}, $\delta\Psi_1$,
$\delta\Psi_0$, and $\delta\Psi_{-1}$ represent the density, spin
and ``quadrupolar" spin fluctuations in the condensate,
respectively, since $\delta
n=\sqrt{n_0}(\delta\Psi_1+\delta\Psi^*_1)$, $\delta M_-
=\sqrt{n_0}\delta\Psi^*_0$ and $\delta M^2_-
=2\sqrt{n_0}\delta\Psi^*_{-1}$. Equations (\ref{flu}) can be
simplified further making use of the conditions under which the
second-order phase transitions occur,
\begin{eqnarray}
  a M_0 + b M_0^3 - g {\Phi_1}^2 &=& 0 , \\
- g M_0 + \alpha + (\beta_0+\beta_s) {\Phi_1}^2 &=& 0 .
\end{eqnarray}

In order to derive dispersion relations of the collective modes,
Equations (\ref{flu}) should be expressed in the momentum space by
performing Fourier transforms, $\delta\Psi_\sigma({\bf
r})=\sum_{\bf k}\delta\Psi_\sigma({\bf k}){\rm exp}(i{\bf k\cdot
r})$, $\delta M_-({\bf r})=\sum_{\bf k}\delta M_-({\bf k}){\rm
exp}(i{\bf k \cdot r})$. For the density fluctuation of the
condensate, we obtain
\begin{eqnarray}
f^I(\delta\Psi^*_1,\delta\Psi_1) &&= {\sum_{\bf k}}^\prime
  \begin{pmatrix}\delta\Psi^*_1({\bf k}) & \delta\Psi_1(-{\bf k})
  \end{pmatrix}
  \times \\
  &&  \begin{pmatrix}\epsilon_k+\beta {\Phi_1}^2 &
      \beta {\Phi_1}^2 \\  \beta {\Phi_1}^2
      & \epsilon_k+\beta {\Phi_1}^2 \end{pmatrix}
    \begin{pmatrix}\delta\Psi_1({\bf k})\\ \delta\Psi^*_1(-{\bf k})
    \end{pmatrix}, \nonumber
\end{eqnarray}
where $\beta=\beta_0+\beta_s$ and ${\sum_{\bf k}}^\prime$ means
the summation is taken only over ${\bf k}>0$. The frequency of
$\delta\Psi_1$ is given by:
\begin{equation}
\hbar \omega_1 = \sqrt{\epsilon_k ^2 + 2\beta{\Phi_1}^2 \epsilon_k
} ,
\end{equation}
which is of the Bogliubov from\cite{ho,ohmi}. Notet here $\beta$
is a phenomenological parameter, not the real interaction
strength.

For the ``quadrupolar" spin fluctuation, we have
\begin{equation}
f^I(\Psi^*_{-1},\Psi_{-1}) = \sum_{\bf k} \delta\Psi^*_{-1}({\bf
k}) (\epsilon_k + 2g M_0-2\beta_s{\Phi_1}^2 ) \delta\Psi_{-1}({\bf
k}) .
\end{equation}
Owing to the magnetization in the normal gas, the gap in the
frequency of $\delta\Psi_{-1}$ becomes larger. So basically, this
kind of excitation is negligible in the Bose ferromagnets.

Our main purpose is to derive spin waves (magnons) in the coexisting
state of the Bose ferromagnet. It can be seen from Eqs. (\ref{flu}) that
{\it the spin waves in the thermal could and in the condensate are
coupled together},
\begin{eqnarray}
f^I(\delta M_+,\delta M_- ; \delta\Psi^*_0,\delta\Psi_0)
  = \sum_{\bf k}  \begin{pmatrix}{\delta M_+({\bf k})} &
  \delta\Psi^*_0({\bf k}) \end{pmatrix} \times \nonumber \\
    \begin{pmatrix}ck^2+\frac {g{\Phi_1}^2}{2M_0} &
      -\frac {\sqrt{2}}2 g \Phi_1 \\  -\frac {\sqrt{2}}2 g \Phi_1
      & \epsilon_k +gM_0  \end{pmatrix}
    \begin{pmatrix}{\delta M_-({\bf k})} \\ \delta\Psi_0({\bf k})
    \end{pmatrix} .
\end{eqnarray}
This equation indicates that the spin fluctuations in each phase
become {\it gapped} solo (see the diagonal matrix elements),
due to the coexistence of the two phases. It seems violating the
Goldstone theorem. To check this, we calculate the
spectrum of the coupled spin mode, $\omega_0$, which is determined by
\begin{eqnarray}\label{spmix}
\begin{vmatrix}ck^2+\frac {g{\Phi_1}^2}{2M_0} - \hbar \omega_0 &
      -\frac {\sqrt{2}}2 g \Phi_1 \\  -\frac {\sqrt{2}}2 g \Phi_1
      & \epsilon_k +gM_0 + \hbar \omega_0 \end{vmatrix}=0 .
\end{eqnarray}
There should be two solutions to Equation (\ref{spmix}), but only one
solution is positive,
\begin{equation}
\hbar \omega_0 = \frac {2\left( ck^2\epsilon_k+\frac
{g{\Phi_1}^2}{2M_0}\epsilon_k+gM_0 c k^2 \right)} {
|\Delta\epsilon|+\sqrt{(\Delta\epsilon)^2+4\left(
ck^2\epsilon_k+\frac {g{\Phi_1}^2}{2M_0}\epsilon_k+gM_0 c k^2
\right)} } ,
\end{equation}
where $\Delta\epsilon = ck^2+\frac {g{\Phi_1}^2}{2M_0} -
\epsilon_k - gM_0$. It implies that {\it only one mixed spin-wave mode
exists inside the Bose ferromagnet} although the system is separated
into ``two fluids", different from the case of sound waves. This is
physically reasonable, since spin wave is associated with the
Goldstone mode and the FM transition happens only once. This mode
mixes spin waves in the ``two fluids".

Next let us look at the long wavelength dispersion of the spin wave.
At very small $k$, the spectrum reduces to
\begin{equation}
\hbar \omega_0 \approx \frac {gM_0ck^2+\frac {g{\Phi_1}^2}{2M_0}\epsilon_k}
           {gM_0-\frac {g{\Phi_1}^2}{2M_0}} .
\end{equation}
Once again, we obtain the $k^2$-formed dispersion relation, and the
Goldstone mode is recovered. We note that the coupling constant $g$
is cancelled in the long wavelength spectrum. Since this result holds
only in the vicinity of the BEC temperature, where the BEC order
parameter is very small, $\Phi_1^2<<M_0$, so the spectrum can be
simplified further to
\begin{equation} \hbar \omega_0 \approx c_s k^2, ~~~{\rm with}~~~
c_s= c+\frac 12 \frac {\Phi_1^2}{M_0^2}\frac {\hbar^2}{2m} .
\end{equation}
$c_s$ is the spin-wave stiffness of the mixed spin-wave mode and
it contains contributions from both the two fluids. But $c_s$ is
not the simple summation of the spin-wave stiffness in the two
phases. A weight factor defined by $\frac 12 \frac
{\Phi_1^2}{M_0^2}$ is applied to the condensate spin-wave
stiffness. Usually $c$ changes smoothly with temperature under
$T_f$. Owing to the spin-wave stiffness of the condensate, there
should be a sharp bend in $c_s$ at $T_c$, which could be measured
experimentally.

In 2002, the JILA group observed spin-state segregation in an
ultracold, noncondensed Rb gas with two sub-spin
states\cite{lewan}. This effect can be interpreted as spin wave in
a pseudo-spin-1/2 Bose gas\cite{oktel}. In a more recent
experiment, similar phenomenon was studied in a partially
Bose-condensed gas\cite{mcguirk}. Coherent coupling between the
normal and superfluid components was observed. Our work is
relevant to the latter case\cite{note4}. As we predict above, the
spin waves in two components should be coupled into one mixed mode
(the coherent mode).

For triplet superconductors, the order parameter of the SC phase
is in a $2\times 2$ matrix form, $\Delta({\bf k})=i[{\bf d}({\bf
k})\cdot {\bf \sigma}]\sigma_y$. The product $\Delta({\bf
k})\Delta({\bf k})^\dag=|{\bf d}({\bf k})|^2\sigma_0+i[{\bf
d}({\bf k}) \times{\bf d}^*({\bf k})] \cdot {\bf \sigma}$, where
${\bf \sigma}=(\sigma_x,\sigma_y,\sigma_z)$ is the Pauli matrix
and ${\bf m}({\bf k})\propto i[{\bf d}({\bf k}) \times{\bf
d}^*({\bf k})]$ acts as magnetic moment of Cooper pairs. It is
suggested that the strong internal field in SC ferromagnets
stabilize a nonunitary triplet state\cite{machida}, which means
that Cooper pairs carry a net magnetization ${\bf m} \ne 0$. This
case is equivalent to the polarization of the FM spinor Bose
condensate, and thus the system consists of two coexisting
ferromagnetic phases. Seeing the similarity between the Bose and
SC ferromagnets, our results for spinor bosons help to understand
spin waves in SC ferromagnets. This problem has drawn
current research interest\cite{braude}. The present
phenomenological approach can be extended to SC ferromagnets.

In summary, we have shown that the Bose ferromagnet consists of
``two fluids" at low temperatures, namely, the magnetized normal
and Bose-condensed phases. Nevertheless, only one mixed spin-wave
mode exists inside. The spin-wave spectrum has the same momentum
dependence as that in a conventional ferromagnet at long
wavelengths, $\omega = c_s k^2$. However, the spin-wave stiffness
$c_s$ contains contributions from the two different phases. $c_s$
is a function of the condensate density $n_0$ below BEC, so it can
exhibit a sharp bend at the BEC temperature, which could be
examined experimentally.

We thank Luis Santos for critical reading of the manuscript and
the Deutsche Forschungsgemeinschaft for support of this work. QG
acknowledge helpful discussions with Kurt Scharnberg and Dimo I.
Uzunov.

%%%%%%%%%%%%%%%%%%%%%%%%%%%%%%%%%%%%%%%%%%%%%%%%%%


\begin{thebibliography}{999}
\bibitem{note2} We note that the free energy density for the FM normal phase can be written in
the form of Equation (1a) when ${\Phi_1^2}<<{M_0}$.
\bibitem{note3} It is approximately shown in Ref. \protect\cite{Gu2} that
$\delta T_f \sim \delta T_c \sim I$ at very small $I$. However more sophisticated treatment is
required to clarify the relation between $\delta T_f$, $\delta T_c$ and $I$.
\bibitem{note4} Our theory is valid in the thermodynamic limit in which the spin
conservation symmetry can be broken spontaneously while in experimental conditions the spin
conservation rule holds so that the FM transition is hindered.
\end{thebibliography}
\end{document}